\documentclass[a4paper,11pt]{article}
\usepackage{pos}
\usepackage{hyperref}
\usepackage{aas_macros}
\usepackage{color}
\newcommand{\C}{3C\,84}

\title{An extensive analysis of the sub-parsec region of 3C\,84}

\author*[a]{G.~F. Paraschos}
\author[e,a]{J.-Y.~Kim}
\author[a]{T.~P.~Krichbaum}
\author[f, c]{J.~Oh}
\author[c, b]{J.~A.~Hodgson}
\author[d]{M.~A.~Gurwell}
\author[a]{J.~A.~Zensus}
\affiliation[a]{$^1$ Max-Planck-Institut f\"ur Radioastronomie,\\ Auf dem H\"ugel 69, Bonn, D-53121 Bonn, Germany\\}
\affiliation[b]{Korea Astronomy and Space Science Institute, 776 Daedeokdae-ro, Yuseong-gu, Daejeon 30455, Korea}
\affiliation[c]{Department of Physics and Astronomy, Sejong University, 209 Neungdong-ro, Gwangjin-gu, Seoul 05006, Korea}
\affiliation[d]{Center for Astrophysics \textbar ~Harvard \& Smithsonian, 60 Garden Street, Cambridge, MA 02138, USA}
\affiliation[e]{Department of Astronomy and Atmospheric Sciences, Kyungpook National University, Daegu 702-701, Republic of Korea}
\affiliation[f]{Joint Institute for VLBI ERIC, Oude Hoogeveensedijk 4, 7991 PD, Dwingeloo, The Netherlands}

\emailAdd{gfparaschos@mpifr-bonn.mpg.de} 

\abstract{The study of jet launching in AGN is an important research method to better understand super-massive black holes (SMBHs) and their immediate surroundings.
The main theoretical jet launching scenarios invoke either magnetic field lines anchored to the black holes's (BH) accretion disc \citep{Blandford82} or a magnetic field, which is directly connected to its rotating ergosphere \citep{Blandford77}.  

The nearby and bright radio galaxy
\C\ (NGC\,1275) is a very suitable target for testing different jet launching mechanisms, as well as for the study of the innermost, sub-parsec scale AGN structure and the jet origin.

Very long baseline interferometry (VLBI) -- specifically at millimetre wavelengths -- offers an unparalleled view into the physical processes in action, in the close vicinity of SMBHs.
Utilising such mm-VLBI observations of \C, we study the jet kinematics of the VLBI core
region of \C\ by employing all available, high sensitivity 3\,mm-VLBI data sets of this source.
As part of this analysis we associate the component ejection events with the variability light-curves at different radio frequencies and in the $\gamma$-rays.
Furthermore, by cross-correlating these light-curves, we determine their time-lags and draw conclusions regarding the location of the high energy emission close to the jet base.
}

\FullConference{%
  15th European VLBI Network Mini-Symposium and Users' Meeting (EVN2022)\\
  11-15 July 2022\\
  University College Cork, Ireland
}

\begin{document}
\maketitle

\section{Kinematics in the radio source 3C\,84}

\begin{figure}
    \centering
    \includegraphics[width=1\columnwidth]{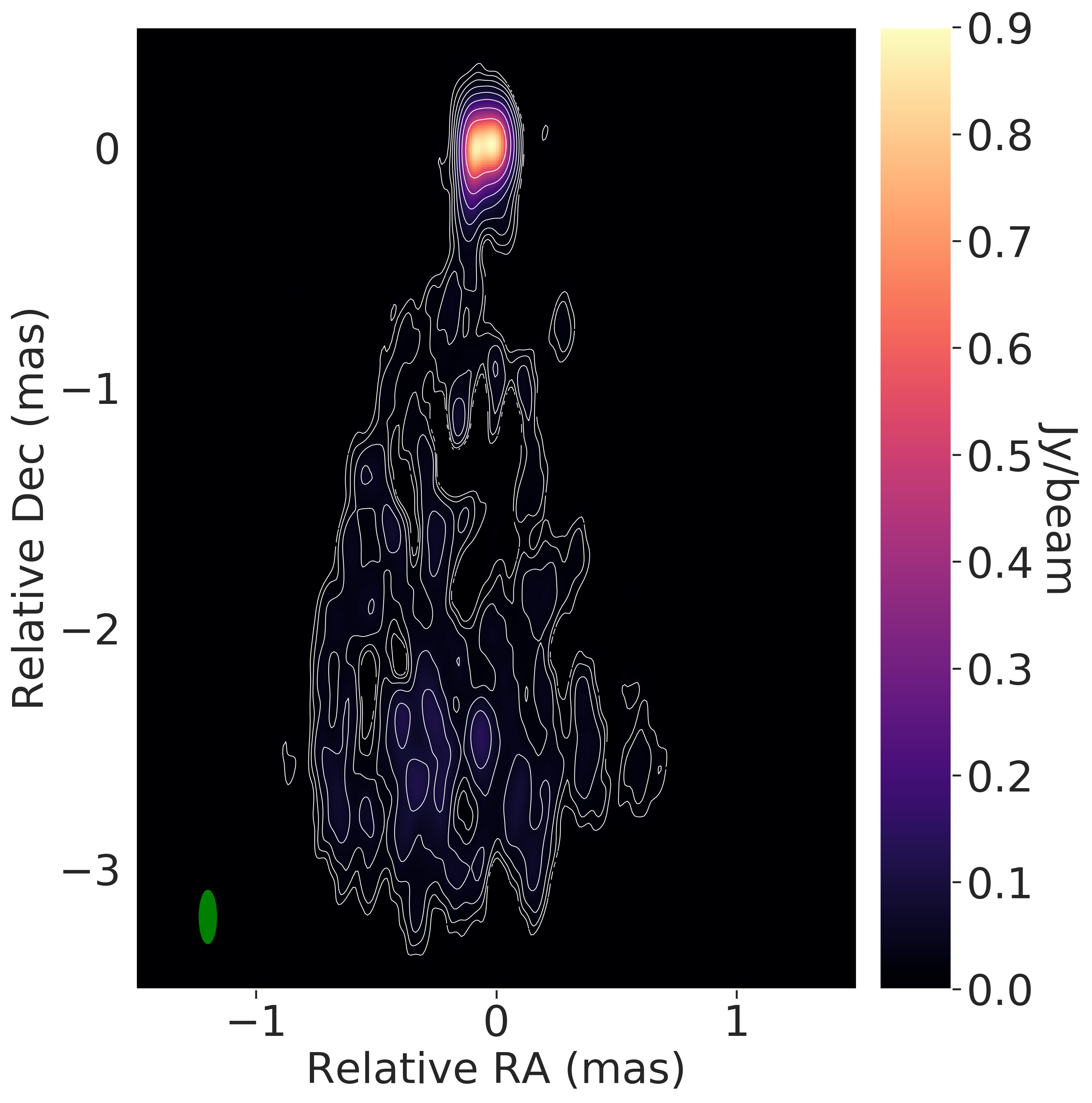}
    \caption{\small \sl
    Stokes I image of \C\ observed with the GMVA at 3\,mm in September 2014.
    The core exhibits an elongated region, which is perpendicularly oriented to the bulk jet flow.
    The contour levels correspond to (0.005, 0.009, 0.02, 0.03, 0.05, 0.09, 0.16, 0.28, 0.49, 0.88) Jy/beam.
    The beam size ($0.22\times0.07$\,mas) is denoted as a green ellipse in the bottom left corner.
    }
    \label{fig:structure}
\end{figure}

The radio source \C\ (NGC\,1275; z=0.0176 \citep{Strauss92}, see Fig.~\ref{fig:structure}) is one of the brightest radio galaxies in the northern sky and a TeV emitter \cite{Magic18}.
The relatively large viewing angle (estimated to be $11^\circ-35^\circ$ \cite{Abdo09, Oh22} in the nuclear region) at which its jet is moving with regard to our line of sight, combined with the subluminal flow speed of its jet features \cite{Krichbaum92, Dhawan98, Hodgson21, Paraschos22a}, allows us to effectively study its jet kinematics, as the Doppler-beaming effect is thus not strong.

In the work, presented in \cite{Paraschos22a}, we used over 20 years worth of Global mm-VLBI Array (GMVA) data for \C\ at 86\,GHz, and, in combination with 43\,GHz data \cite{Jorstad05, Jorstad17}, we were able to cross-identify at both frequencies and track inside the nuclear region, five jet features, named $\textrm{F}_1-\textrm{F}_5$.
We fitted a linear regression to the VLBI component trajectories, to calculate their velocity and also back-extrapolate their time of ejection.
Our work reveals that the features move with velocities up to $\sim0.1$c in the nuclear region, that newer components (after 2016) seem to move faster, and that at 86\,GHz, the features move at approximately twice the speed as at 43\,GHz.
%

\section{Total flux density variability and feature ejection association} \label{sect:variability}

Oftentimes, the emergence of VLBI features in AGN-jets can be associated with flares in the radio-bands and in the $\gamma$-rays \cite{Savolainen02, Karamanavis16}.
To test if this is also the case in \C, we compared the temporal point of ejection of the features $\textrm{F}_1-\textrm{F}_5$ with available radio- and $\gamma$-ray light curves (see also Fig.~4 in \cite{Paraschos22a}).
In our cross-correlation analysis we employed radio light curves at 4.8, 8.0, and 14.8\,GHz, obtained at University of Michigan Radio Observatory (UMRAO), at 15\,GHz obtained at the Owen’s Valley Radio Observatory (OVRO; \cite{Richards11}), at 37\,GHz obtained at the Metsähovi Radio Observatory (MRO), at 230 and 345\,GHz obtained with the Submillimeter Array (SMA), and the $\gamma$-ray light curve of NGC\,1275 (\C) at MeV-GeV energies \cite{Atwood09, Kocevski21}.
For a more detailed description refer to \cite{Paraschos22a}.

We do not find a clear relation between the ejection times of jet features and flaring activity in both radio- and $\gamma$-ray light curves.
$\textrm{F}_1$ and $\textrm{F}_2$ do not seem to be associated with any activity in the available centimetre radio flux \cite{Paraschos22a}.
$\textrm{F}_3$, $\textrm{F}_4$, and $\textrm{F}_5$ on the other hand seem to be associated with radio- and $\gamma$-ray flares but it is not always clear if the radio flux is preceding or trailing the $\gamma$-ray flux (see top of Fig. \ref{fig:flares}). Our work is presented in more detail in \cite{Paraschos22b}.

\section{Light curve cross-correlation analysis}
\subsection{Core shift}

\begin{figure}
    \centering
    \includegraphics[width=1\columnwidth]{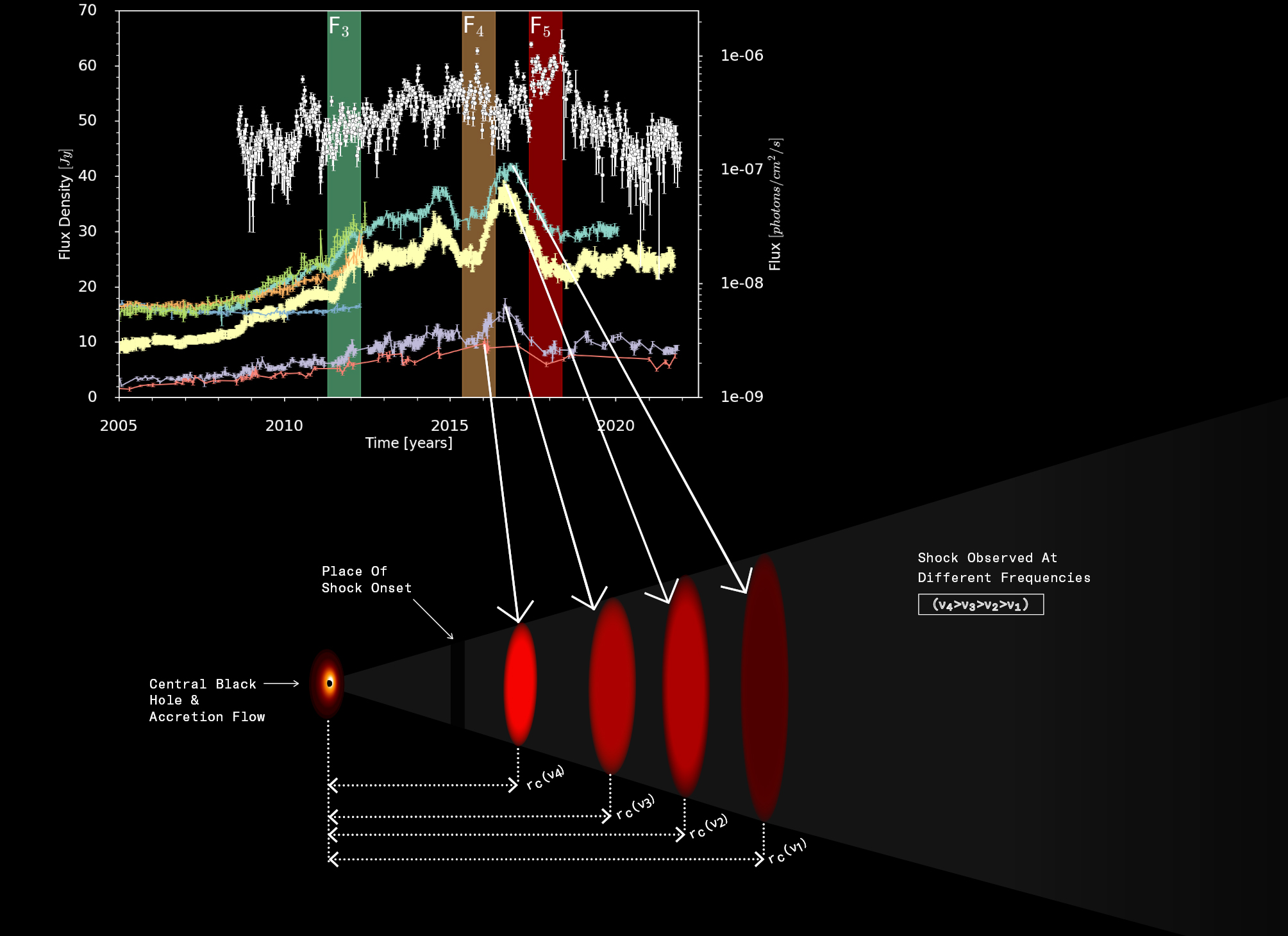}
    \caption{\small \sl 
    Illustration of the appearance of shocks in a conical and homogeneous jet \cite{Blandford79} depending on the observing frequency ($\nu_1-\nu_4$).
    The top panel displays (from top to bottom) the light curves at $\gamma$-rays (white), 15\,GHz (light blue), 37\,GHz (yellow) 14.8\,GHz (green), 8.0\,GHz (orange), 4.8\,GHz (dark blue), 230\,GHz (purple), and 345\,GHz (red).
    The 2016 flare is first visible in the millimetre radio flux and starts getting visible later on downstream in the centimetre radio flux.
    Arrows are used to guide the eye.
    The shaded regions denote the temporal points of ejection of the features $\textrm{F}_3$, $\textrm{F}_4$, and $\textrm{F}_5$.
    The apparent distance of the shock onset ($r_\mathrm{c}(\nu)$) from the black hole is denoted with the orange ellipses in the bottom panel.
    Figure inspired by Fig.~1 in \cite{Kudryavtseva11}.}
    \label{fig:flares}
\end{figure}

In VLBI images of jets, a compact, non-expanding, bright region, which exhibits a flat spectrum is called the core.
The core is thought to be characterised by an optically thick surface \cite{Kudryavtseva11} and the position of it is thus dependent on the observing frequency (assuming a conical, homogeneous jet, as, for example, described in \cite{Blandford79, Konigl81}).
This opacity effect, which shifts the apparent location of the core in the sky, is called the core shift \cite{Lobanov98b}.
The core shift can be estimated using total variability light curves, as explained, for example in \cite{Kudryavtseva11, Karamanavis16, Kutkin18} (see also Fig.~\ref{fig:flares}).
For our analysis we used the light curves described in Sect.~\ref{sect:variability} and also added a 91.5\,GHz light curve obtained by the Atacama Large Millimeter Array (ALMA)\footnote{Details regarding the estimation of the flux are presented in the \href{https://almascience.eso.org/documents-and-tools/cycle8/flux-service-of-the-alma-source-catalogue}{Flux Service of the ALMA Source Catalogue} manual.}.

We used two approaches to cross-correlate the available light curves: 1) the discrete cross-correlation function (DCF) \cite{Edelson88}, which is a procedure used to measure correlations between data sets with known measurement errors, which can even be unevenly sampled (see also \cite{Hovatta07, Kutkin14, Fuhrmann14, Rani17, Hodgson18}); and the so-called 2) Gaussian process regression (GPR) \cite{Rasmussen06}, which is a machine learning procedure that utilises a non-parametric approach to data fitting, without prerequisite assumptions about the fitting function (see also \cite{Karamanavis16, Kutkin18, Mertens18, Pushkarev19}).
A detailed discussion about both methods is presented in \cite{Paraschos22b}.

We utilised both, the GPR and the DCF method to cross-correlate the radio light curves at each frequency, to its neighbouring frequency (for example 4.8\,GHz with 8.0\,GHz, 8.0\,GHz with 15\,GHz, etc.) and from this we calculated the time lags, that is, the time it takes for a flare at frequency $\nu_2$ to appear at frequency $\nu_1$ ($\nu_2>\nu_1$).
We then fitted a power law of the form $\Delta t\propto \nu^{1/k_\textrm{r}}$ to the averaged time lags acquired with both methods.
The fit yielded $k_\textrm{r} = 1.08\pm0.18$. 
In order to then transform the core shift (see \cite{Lobanov98b, Hirotani05} for a detailed description) in physical units of distance, we adopted average component velocities in the nuclear region, in the range of $\beta_\textrm{app}=0.03-0.12$c.
This computation yields distances in the range of $z_\textrm{3\,mm} = 22 - 645$ Schwarzschild radii ($R_\textrm{s}$) between the jet base and the 3\,mm VLBI core.
This range is also in agreement with another core shift estimate presented in \cite{Paraschos21, Oh22}.
The interested reader is referred to \cite{Paraschos22b} for a more in-depth presentation of the analysis.



\subsection{Location of $\gamma$-ray emission}

An open question regarding the jet in \C\ is the location of the $\gamma$-ray emission within the jet (see, e.g. \cite{Nagai12}). 
Both an upstream \cite{Ramakrishnan15} and a downstream location \cite{Hodgson18, Britzen19} of the $\gamma$-ray emission site, with regard to the radio emission site, have been suggested.

\begin{figure}
    \centering
    \includegraphics[width=1\columnwidth]{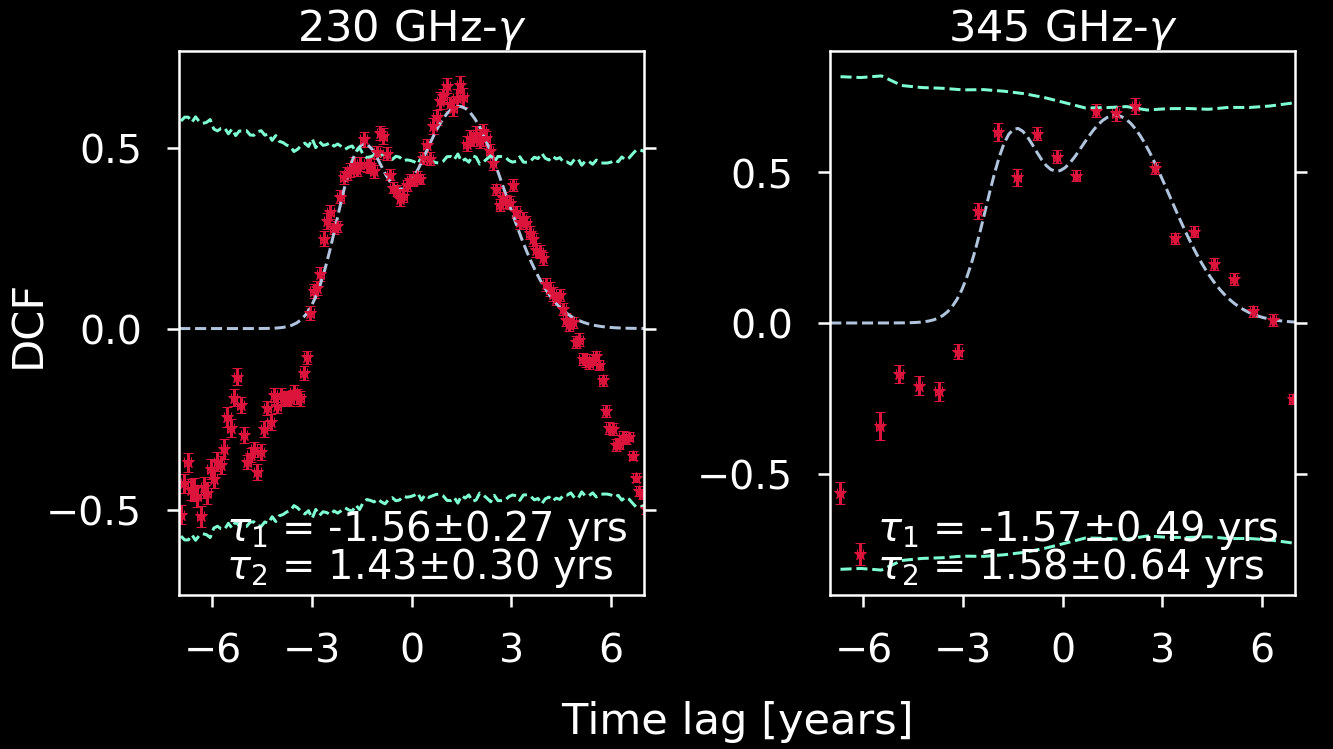}
    \caption{\small \sl
    DCF analysis of $\gamma$-ray light curve. Left: DCF of $\gamma$-rays versus 230\,GHz for \C. 
    A double Gaussian function (dashed light-blue line) was fit to the DCF, to estimate more accurately the two peak positions (using the means $\tau_1$ and $\tau_2$).
    The dashed green curve denotes the 99.7\% confidence band.
    Right: Same for 345\,GHz versus $\gamma$-rays.
    }
    \label{fig:gamma-radio}
\end{figure}

Using the DCF approach (see Fig.~\ref{fig:gamma-radio}), we find that the $\gamma$-rays may either precede the 230\,GHz flux 
by $\tau_{\gamma-230\,\textrm{GHz}} = 1.56 \pm 0.27$ years 
or trail the 230\,GHz flux 
by $\tau_{\textrm{230\,GHz}-\gamma} = 1.43 \pm 0.30$ years 
Since we find positive and negative time lags between the radio emission and the $\gamma$-rays (see \cite{Paraschos22b} for a detailed presentation and discussion of the results), this might indicate multiple locations of emission of the $\gamma$-rays, both upstream and downstream of the radio emission, depending on individual flares.
At 345\,GHz we see a similar effect. However, the sparse time sampling of the 345 GHz light curve formally limits the significance of the cross-correlation.
Ideas to explain the $\gamma$-ray emission invoke an origin
in the parsec-scale jet and multi-zone emission \cite{Marscher14, Hodgson18},
or `mini-jets' \cite{Giannios13, Hodgson21}.

\section{Conclusions}

We have presented a high-resolution kinematics study of the jet of \C, using millimetre-VLBI observations, and a search for possible correlations with flux density variability in the radio- and $\gamma$-ray bands.
We find:
\begin{enumerate}
    \item Jet features move subluminally in the nuclear region of \C\ (with apparent speeds up to $0.1$c), with newer ejected features moving faster.
    We find evidence of faster motion (by a factor of two) at 86\,GHz when compared to 43\,GHz.
    \item A clear association between VLBI component ejection and flaring activity is not found. 
    We note that jet component $\textrm{F}_4$ was ejected at the onset of a radio flare, but component $\textrm{F}_5$ during a phase of declining radio flux.
    \item The jet apex of \C\ is located at $z_\textrm{3\,mm} = 22 - 645\,R_\textrm{s}$ upstream of 3\,mm VLBI core, which compares well to previous work \cite{Paraschos21, Oh22}.
    \item The presence of two correlation peaks in the DCF plot of the radio- and $\gamma$-ray flux suggests multiple locations of the $\gamma$-ray emission region further downstream in the bulk jet flow.
\end{enumerate}

\acknowledgments
\noindent
This work makes use of light curves at the following frequencies: 4.8, 8.0, and 14.8\,GHz, kindly provided by the University of Michigan Radio Astronomy Observatory which has been supported by the University of Michigan and by a series of grants from the National Science Foundation, most recently AST-0607523;
15\,GHz kindly provided by the OVRO 40-m monitoring program \cite{Richards11}, supported by private funding from the California Institute of Technology and the Max Planck Institute for Radio Astronomy, and by NASA grants NNX08AW31G, NNX11A043G, and NNX14AQ89G and NSF grants AST-0808050 and AST-1109911.; 
37\,GHz kindly provided by the Aalto University Metsähovi Radio Observatory, 
91.5\,GHz taken from the ALMA Calibrator source catalogue; and 230\,GHz and 345\,GHz kindly provided by the Submillimeter Array, respectively.
Maunakea is a culturally important site for the indigenous Hawaiian people; we are privileged to study the cosmos from its summit.
G. F. Paraschos is supported for this research by the International Max-Planck Research School (IMPRS) for Astronomy and Astrophysics at the University of Bonn and Cologne. This research has made use of data obtained with the Global Millimeter VLBI Array (GMVA), which consists of telescopes operated by the MPIfR, IRAM, Onsala, Metsähovi, Yebes, the Korean VLBI Network, the Green Bank Observatory and the VLBA (NRAO). The Submillimeter Array (SMA) is a joint project between the Smithsonian Astrophysical Observatory and the Academia Sinica Institute of Astronomy and Astrophysics and is funded by the Smithsonian Institution and the Academia Sinica. This research has also made use of data from the Owen's Valley Radio Observatory 40-m monitoring program \citep{Richards11}, supported by private funding from the California Insitute of Technology and the Max Planck Institute for Radio Astronomy, and by NASA grants NNX08AW31G, NNX11A043G, and NNX14AQ89G and NSF grants AST-0808050 and AST-1109911.
Finally, this research makes use the publicly available $\gamma$-ray light curve of NGC\,1275 (\url{https://fermi.gsfc.nasa.gov/ssc/data/access/lat/msl_lc/source/NGC_1275}).

\bibliographystyle{JHEP}
\bibliography{skeleton}

\end{document}